\documentclass[a4paper,11pt]{article}
\pdfoutput=1 

\usepackage{jinstpub} 

\usepackage{lineno}
\usepackage{url}

\proceeding{Topical Workshop on Electronics for Particle Physics - TWEPP2025\\
	Oct 06-10, 2025\\
	Rethymno, Crete, Greece}

\title{Advancements and future expansions of the Caribou DAQ system}

\author[a, 1]{Y. Otarid,\note{Corresponding author.}}
\author[b]{M. Benoit,}
\author[c]{E. Buschmann,}
\author[c]{H. Chen,}
\author[a]{D. Dannheim,}
\author[a]{I. Kamoisis,}
\author[e]{\mbox{T. Koffas,}}
\author[e]{\mbox{R. St-Jean,}}
\author[d]{S. Spannagel,}
\author[c]{S. Tang,}
\author[d]{T. Vanat,}
\author[e]{C. You}

\affiliation[a]{CERN,\\Esplanade des Particules, Geneva ,Switzerland}
\affiliation[b]{ORNL,\\1 Bethel Valley Rd, Oak Ridge, United States}
\affiliation[c]{BNL,\\98 Rochester St, Upton, United States}
\affiliation[d]{DESY,\\Notkestrasse 85, Hamburg, Germany}
\affiliation[e]{Carleton University,\\1125 Colonel By Dr, Ottawa, Canada}

\emailAdd{younes.otarid@cern.ch}

\abstract{Caribou is a versatile data acquisition (DAQ) system developed within several collaborative frameworks (CERN EP R\&D, DRD3, AIDAinnova, and Tangerine) to support laboratory and test-beam characterization of novel silicon pixel detectors. It combines a custom Control and Readout (CaR) board with a Xilinx Zynq System-on-Chip (SoC) running project-wide shared firmware and software stacks. The system architecture emphasizes reusability, flexibility, and ease of integration. The CaR board provides essential interfaces such as programmable power supplies, voltage and current references, high-speed ADCs, and configurable I/O lines for detector control and readout. The SoC runs an embedded Linux distribution built with PetaLinux and integrates two main components: Peary, a C++ embedded DAQ application providing hardware abstraction, configuration management, logging, and multi-device control through Command Line (CLI) and Python interfaces; and Boreal, a common Caribou FPGA firmware framework offering reusable modules and automated build workflows for user-specific bit files. The next major milestone in Caribou’s evolution is the transition to version 2.0, based on a Zynq UltraScale+ System-on-Module (SoM) architecture. This paper presents the recent progress, and future prospects of the project, and describes recent hardware, firmware, and software developments preparing the system for the upcoming CaR board v2.0.}


\begin{document}
\maketitle
\flushbottom

\section{Introduction}
\label{sec:introduction}

\begin{figure}[h]
	\centering
	\includegraphics[width=\textwidth]{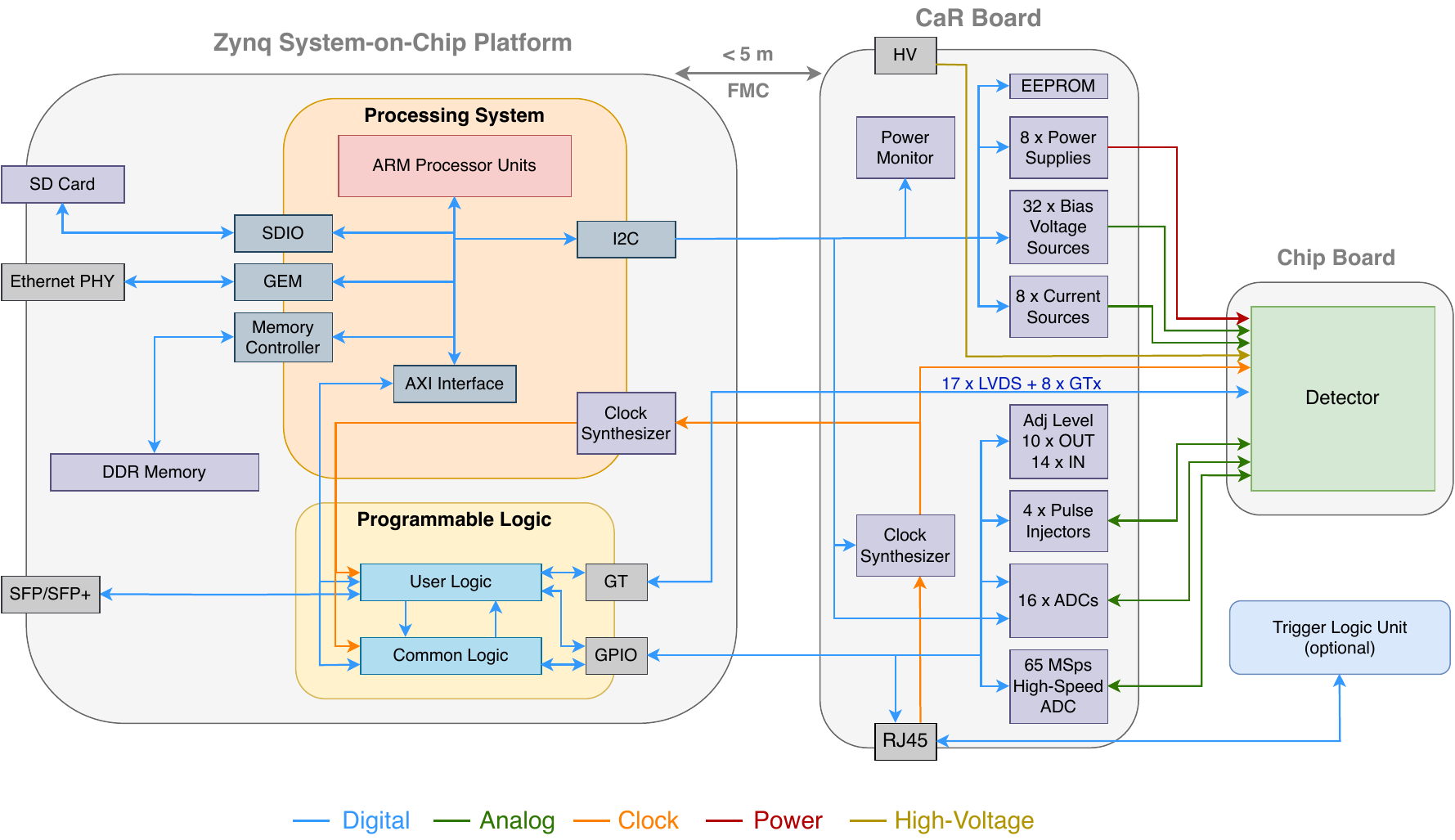} 
	\caption{Block diagram of the Caribou hardware architecture, illustrating all analog and digital components, as well as the system’s available connections and interfaces}
	\label{fig:caribou_architecture}
\end{figure}

The development of new silicon pixel detectors for high-energy physics experiments requires versatile and reusable data acquisition (DAQ) systems capable of supporting rapid prototyping, laboratory qualification, and test-beam operation. The Caribou system \cite{Benoit,Fiergolski,Vanat,Buschmann,Otarid,Dannheim, Caribou Website} was developed to meet these needs within several collaborative frameworks — CERN EP R\&D, DRD3, AIDAinnova, and Tangerine — and is now used by more than fourteen institutes for the characterization of a wide range of detector prototypes. It combines a custom Control and Readout (CaR) board  \cite{Carboard Gitlab} with a Xilinx Zynq System-on-Chip (SoC) platform running a common firmware and software stack shared across projects. Its architecture emphasizes modularity, reusability, and ease of integration, enabling more than fifteen different detector prototypes to be tested within a unified environment. The CaR board provides key interfaces such as programmable power supplies, voltage and current references, high-speed ADCs, injection pulsers, and configurable I/O lines for detector control and readout. It connects to the SoC through FMC and to detector-specific chip boards via a SEARAY connector, allowing adaptation to various detector designs while maintaining a common DAQ backbone. On the software and firmware side, Caribou integrates two central components via its PetaLinux image builder \cite{Peta-Caribou Gitlab} : Peary \cite{Peary Gitlab}, a C++ framework offering hardware abstraction, configuration management, and multi-device control via command-line and Python interfaces; and Boreal \cite{Boreal Gitlab,Boreal Modules Gitlab}, an FPGA firmware framework providing reusable IP cores and automated synthesis and simulation workflows. The use of Vivado Block Design facilitates the coexistence of shared control infrastructure and user-specific logic, while a strict separation between common and detector-specific components ensures maintainability and scalability. Figure~\ref{fig:caribou_architecture} shows a block diagram of the Caribou system hardware architecture. It outlines the key resources provided by the Zynq SoC platform, the CaR board, and the chip board, as well as their interconnections and corresponding signal types.

\section{Advancements and Future Expansions}
\label{sec:advancements_expansions}

Over the past development cycle, significant progress has been achieved across multiple areas of the project. This includes advancements in hardware validation procedures, the hardware design of the upgraded v2 platform, the FPGA firmware infrastructure, and the embedded software stack. Together, these developments aim to strengthen the overall system’s functionality, reliability, and integration capabilities.

\subsection{Hardware Design and Validation}
\label{subsec:hardware}

\begin{figure}[htbp]
	\centering
	\begin{subfigure}[t]{0.64\textwidth}
		\centering
		\includegraphics[width=\linewidth]{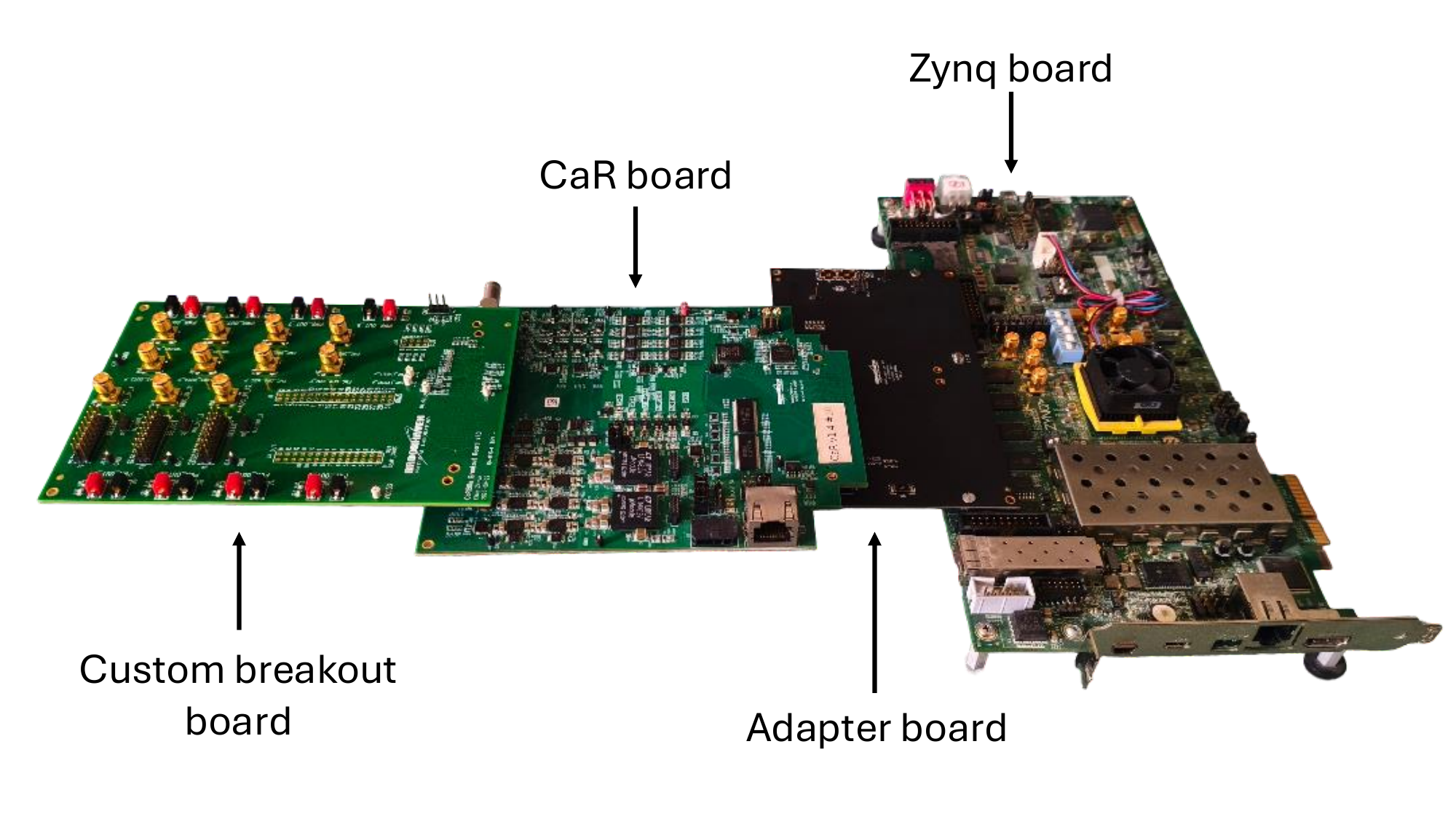}
		\caption{CaR board v1 validation setup, consisting of a Zynq board, the CaR board under test, and a breakout board that provides access to all components and signals available on the CaR board}
		\label{fig:carboard_v1.5_validation_setup}
	\end{subfigure}
	\hfill
	\begin{subfigure}[t]{0.33\textwidth}
		\centering
		\includegraphics[width=\linewidth]{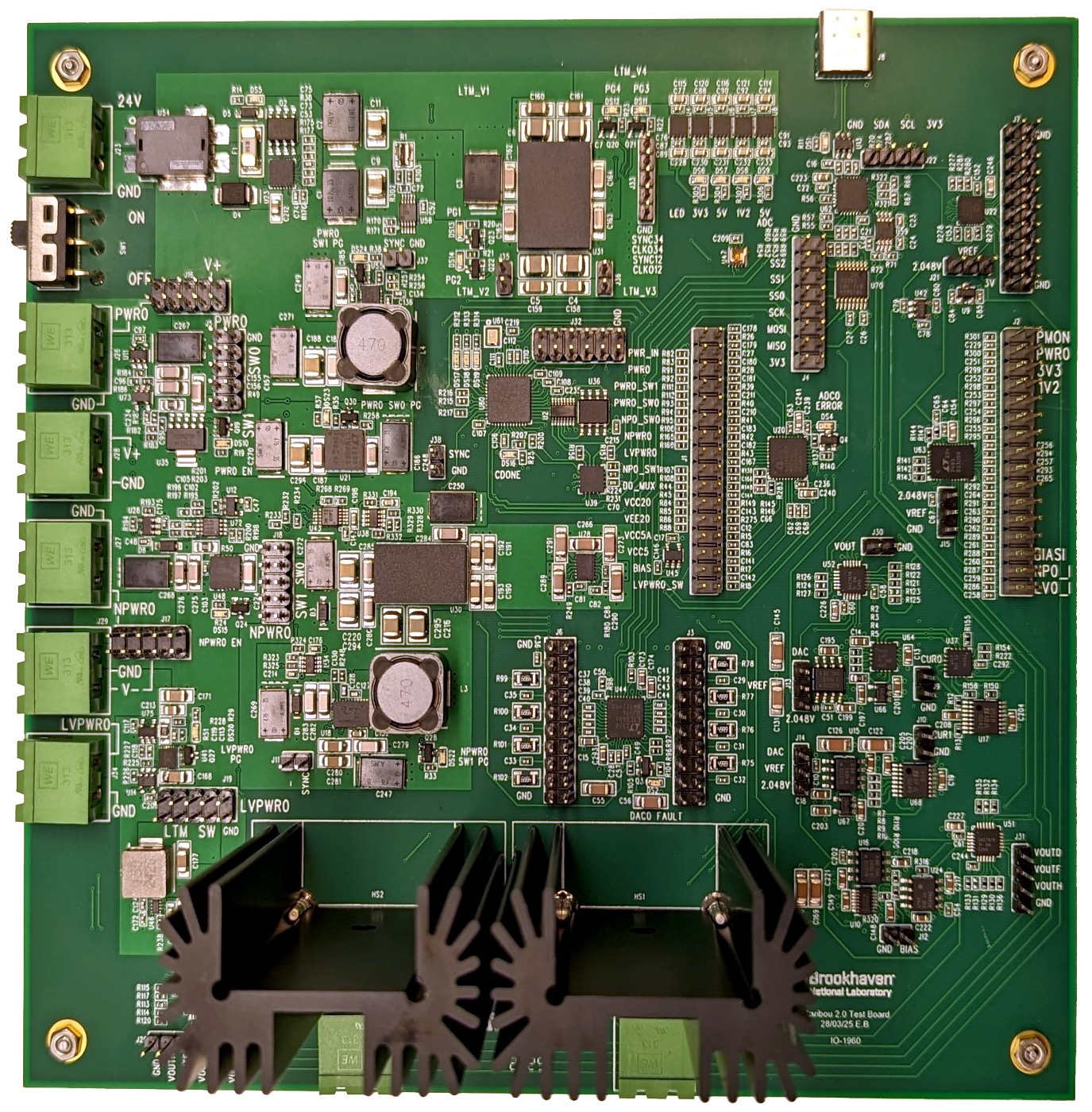}
		\caption{Test board implementing the complete set of features selected for the v2 CaR board}
		\label{fig:carboard_v2_testboard}
	\end{subfigure}
	\caption{Pictures of the CaR board v1 validation setup (a) and the v2 test board (b)}
	\label{fig:hardware_design_validation}
\end{figure}

To validate the functionality of the CaR board v1, a dedicated validation setup was developed. Figure \ref{fig:carboard_v1.5_validation_setup} shows a picture of the setup, which comprises a Zynq board, an adapter board, the CaR board under test, and a custom breakout board. The Zynq board executes the software and firmware routines used to verify the different functionalities of the CaR board. The adapter board serves as an I/O expander, providing access to all CaR board interfaces by utilizing both FMC connectors on the Zynq board. Finally, the custom breakout board enables loopback routing of LVDS signals, includes SMA connectors for high-speed transceiver lines, and provides voltage and current inputs and outputs for both the slow and fast ADCs. The primary objective of this setup is to validate all functionalities of the CaR board by testing its individual components and I/Os upon the arrival of newly produced boards and before their distribution to users. The currently available routines enable verification of the LVDS lines, high-speed transceiver links, power supplies, and ADCs, and are developed within the Boreal \cite{Boreal Gitlab} and Peary \cite{Peary Gitlab} frameworks.\\

A major upgrade of the Caribou system, designated Caribou v2.0, is currently under development to transition the platform to an UltraScale+ System-on-Module architecture with a fully redesigned CaR board. While this evolution primarily targets the system-level integration and SoC platform, it also introduces substantial enhancements to the analog control and monitoring capabilities compared to the existing v1 implementation. The currently deployed CaR board (v1) will be significantly extended in the upcoming v2 revision, which is still under design. While v1 relies primarily on 12-bit DACs for power, bias, and current control, v2 will retain 12-bit resolution for most power rails but extend the available voltage ranges, in particular for higher-voltage supplies. The current source architecture will be substantially improved in v2, introducing multiple selectable bipolar ranges spanning from tens of microamperes with step sizes down to $10 nA$ up to around $100 mA$ with step sizes up to $50 uA$ and a continuous positive/negative range, compared to the limited $1.024 mA$ range of v1. Bias generation will be upgraded from a 12-bit unipolar DAC to a 16-bit multi-range DAC supporting both unipolar and bipolar outputs up to $\pm15 V$, and complemented by a dedicated bias source with programmable current and voltage limits. On the readout side, v2 will replace the low-speed monitoring ADC of v1 with a high-resolution 24-bit ADC optimized for precision measurements, while the $65 MSps$ high-speed ADC present on v1 will be removed. Instead,  v2 will integrate an FMC connector to support compatible expansion modules, such as high-speed ADC FMC cards. Finally, it will incorporate enhanced overcurrent protection implemented on an FPGA.\\

In preparation for the design of the CaR board v2, a simplified test board as shown in Figure \ref{fig:carboard_v2_testboard} was developed. This $16.5\times17.5\ \mathrm{cm^{2}}$ compact board serves as a platform to evaluate the different v2 design options and to validate the performance of previously described target components and circuits, all accessible and controlled through a single USB-to-I$\mathrm{^{2}}$C/SPI interface. The System-on-Module is not included at this stage. Two test boards have been produced and are fully operational, with extensive validation currently in progress. The high-voltage power supplies provide adjustable outputs from approximately $+2.5 V$ to $+18 V$ and $-2.5 V$ to $-18 V$ at up to $1 A$, with one pair intended for the Device Under Test (DUT) and two pairs for powering the DACs, current sources, and bias circuitry. Initial DC-DC converter tests are largely complete and confirm operation up to $17-18 V$ to meet DAC headroom requirements, although further tuning and analysis of regulation, efficiency, and LDO response speed are still ongoing. The low-voltage supplies provide outputs from approximately $0.8 V$ up to $5 V$ at $2 A$, and have been successfully tested down to $2 V$. All tested current sources (two per board on two boards) are operational and show a gain error performance within $\pm0.1-0.25\%$ of the full-scale range. The bias source provides a $\pm15 V$ output range (extendable with a larger supply range) and features a programmable current limit of $\pm10.24 mA$. The DAC provides a $\pm15 V$ output range with up to $55 mA$ per channel, with the maximum current primarily limited by thermal considerations. One of the monitoring ADCs supports a specified input range of $\pm10 V$ and remains functional up to $\pm20 V$. Finally, the current protection circuit and associated FPGA logic were tested and operate as intended: the current can be read back, the current limit and response time are configurable, and channels can be grouped such that a fault in one channel disables all channels in the group.

\subsection{Common FPGA Firmware Modules Design}
\label{subsec:firmware}

To centralize the development of common Caribou soft IPs within the Boreal framework \cite{Boreal Gitlab}, the \mbox{Boreal Modules} \cite{Boreal Modules Gitlab} project was created. Its objective is to provide a unified repository for the design of custom IP cores, supported by a streamlined and flexible verification workflow. No specific simulator is enforced, allowing developers to use the toolchain of their choice; however, the preferred environment currently centers around Cocotb \cite{Cocotb}, the coroutine-based co-simulation framework. At present, the Boreal Modules project includes a variety of soft IPs such as counters, edge detectors, generic AXI4-Lite register maps, and synchronization logic, with continuous expansion to support additional functionalities. Beyond these logic IPs, the project also aims to provide specialized control and readout interfaces for user integration, such as a Trigger Logic Unit (TLU) interface, a CaR board High-Speed ADC interface, and a Time-to-Digital Converter (TDC) block implementation. While several of these components are already supported on the ZC706 platform, ongoing work focuses on migrating and adapting their designs for compatibility with newer UltraScale+ platforms, including the ZCU102, Mercury+ with XU1 SoM, and the upcoming \mbox{CaR board v2.0}.

\subsection{Embedded Software Architecture Design Revision}

With the CaR board v2.0 under development and the portfolio of supported Zynq platforms expanding, a complete redesign of the Peary embedded \mbox{software \cite{Peary Gitlab}} became necessary. This revision focused on introducing a new embedded architecture featuring a modular Hardware Abstraction Layer (HAL) and improving cross-platform support. The goal was to evolve Peary from a ZC706-centered implementation into a flexible framework capable of supporting any combination of CaR board and Zynq platform.

\begin{figure}[h]
	\centering
	\includegraphics[width=0.7\textwidth]{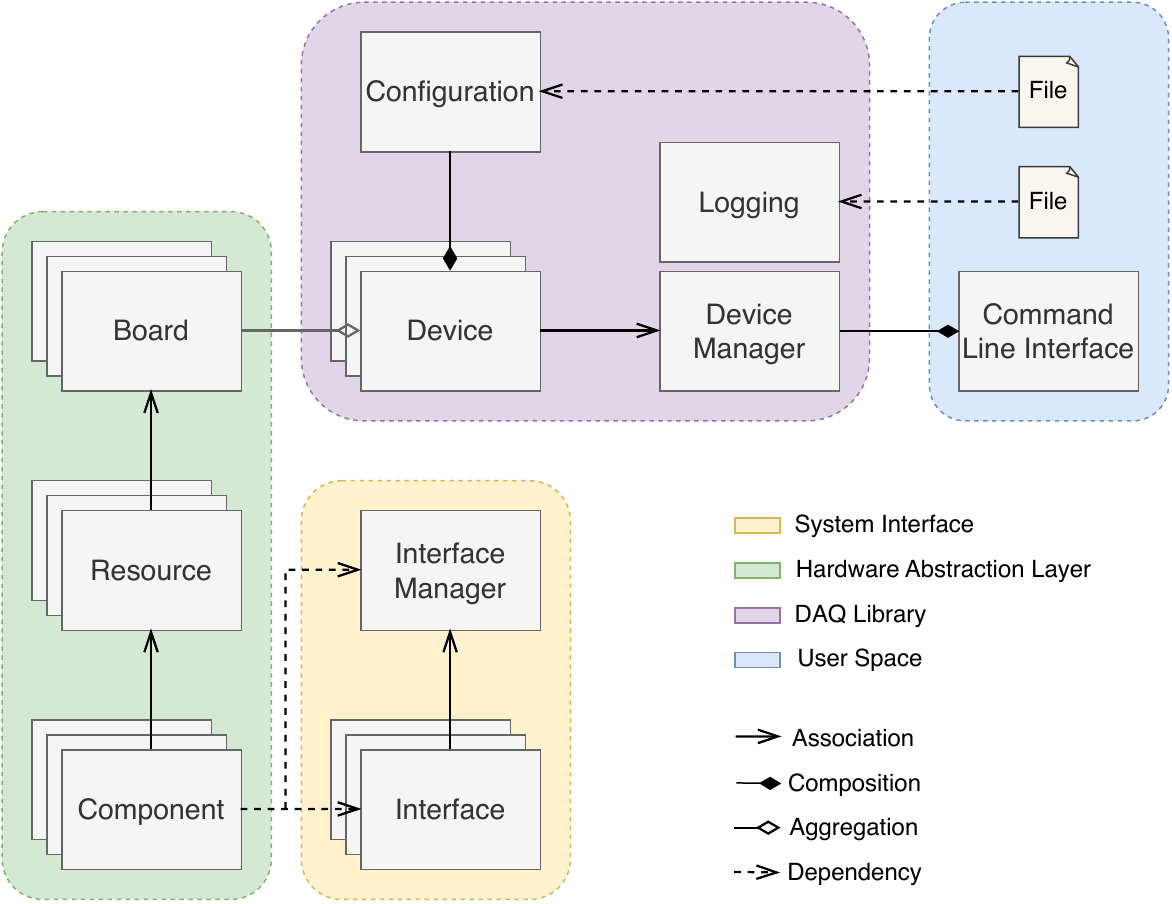} 
	\caption{Simplified UML diagram of the revised Peary architecture, illustrating the structure of the embedded application and the relationships among all system components}
	\label{fig:peary_architecture}
\end{figure}
\label{subsec:software}

As illustrated in Figure \ref{fig:peary_architecture}, the new architecture is organized into four functional spaces. At its core lies the HAL, which is built around the concept of \textit{Resources}. These represent logical board functionalities that may be implemented using one or more hardware components to achieve a specific purpose. For example, supplying a voltage on a given CaR board channel involves several coordinated \textit{Resources} — one to set the output voltage, others to monitor voltage, current, and power consumption, and another to handle overcurrent alerts. Each \textit{Resource} can thus comprise multiple components, and a collection of such \textit{Resources} defines a complete board.
The System Interface space includes all detector and peripheral communication interfaces, such as I²C, SPI, and custom serial links. An Interface Manager oversees access to these interfaces, ensuring safe and coordinated communication without race conditions or interruptions.
The DAQ Library space provides higher-level services such as configuration management (mapping devices under test to setting–value pairs), logging, and device management. It defines a generic Device structure for detector integration and a Device Manager responsible for handling the lifecycle of connected devices.
Finally, a Command Line Interface (CLI) is exposed to users, providing access to all available Peary functionalities and enabling the creation of custom, automated detector qualification routines.
These architectural changes enhance the overall modularity of the system, provide greater flexibility for supporting future boards and Zynq platforms, and enable a more streamlined and scalable detector integration workflow. The new architecture has been successfully tested and will soon be adopted as the baseline for future users.

\section{Conclusion}
The developments presented in this paper mark a significant step forward in the evolution of the Caribou system. On the hardware side, the validation of the CaR board v1 and the design of the CaR board v2 test board have established a robust foundation for the upcoming CaR board v2. These efforts have enabled early testing and optimization of key analog and power-supply circuits, ensuring improved reliability and performance in the next generation of the platform.
In parallel, the complete redesign of the Peary embedded software has introduced a modular and extensible architecture that can support multiple Zynq platforms and future hardware revisions. The introduction of the Hardware Abstraction Layer and the concept of \textit{Resources} provides a flexible framework for integrating new boards and detector devices with minimal reconfiguration. Together, these advancements considerably strengthen the modularity, scalability, and maintainability of the Caribou system.

\section{Acknowledgments}
The developments presented in this contribution are performed within the DRD3 collaboration on Solid State Detectors and in collaboration with the CERN EP R\&D programme on technologies for future experiments. This project has received funding from the European Union’s Horizon 2020 Research and Innovation programme under GA no 101004761. The production of the common hardware is supported by the RD50 and DRD3 Common Funds.


%
%
%
%



\begin{thebibliography}{99}

\bibitem{Benoit}
H. Liu et al, \emph{Development of a modular test system for the silicon sensor R\&D of the ATLAS
	Upgrade}, JINST 12 P01008 (2016)

\bibitem{Fiergolski}
A. Fiergolski, \emph{A multi-chip data acquisition system based on a heterogeneous system-on-chip
	platform}, Springer Proc. Phys TIPP2017 (2017)

\bibitem{Vanat}
T. Vanat et al., \emph{Caribou — A versatile data acquisition system}, PoS, TWEPP2019 (2020)

\bibitem{Buschmann}
E. Buschmann et al., \emph{Status and recent extensions of the Caribou DAQ system for picosecond timing with an FPGA TDC}, JINST 18 C02005, TWEPP2022 (2022)

\bibitem{Otarid}
Y. Otarid et al., \emph{Caribou - A versatile data acquisition system for silicon pixel detector prototyping}, JINST, PIXEL2024 (2025)

\bibitem{Dannheim}
CLICdp Collaboration (eds. D. Dannheim et al), \emph{Detector technologies for CLIC}, CERN Yellow Reports: Monographs, CERN-2019-001 (2019)

\bibitem{Caribou Website}
Caribou project website, \url{https://caribou-project.docs.cern.ch}

\bibitem{Carboard Gitlab}
CaR board GitLab repository, \url{https://gitlab.cern.ch/Caribou/hardware/carboard}

\bibitem{Peta-Caribou Gitlab}
Peta-Caribou Gitlab repository, \url{https://gitlab.cern.ch/Caribou/peta-caribou}

\bibitem{Peary Gitlab}
Peary software GitLab repository, \url{https://gitlab.cern.ch/Caribou/peary}

\bibitem{Boreal Gitlab}
Boreal firmware GitLab repository, \url{https://gitlab.cern.ch/Caribou/boreal}

\bibitem{Boreal Modules Gitlab}
Boreal modules GitLab repository, \url{https://gitlab.cern.ch/Caribou/boreal-modules}

\bibitem{Cocotb}
Cocotb website, \url{https://www.cocotb.org}


%





\end{thebibliography}
\end{document}